\documentclass[aps,11pt,axodraw]{revtex4}
\usepackage{epsfig}
\usepackage{amsmath}
\usepackage{bm}
\usepackage{times}
\usepackage{graphicx}
\usepackage{axodraw}
\usepackage{color}
\begin{document}
\title{\Large ON THE ORIGIN OF NEUTRINO MASSES}
\author{Pavel Fileviez P\'erez$^{1}$}
\author{Mark B. Wise$^{2}$}
\affiliation{$^{1}$University of Wisconsin-Madison, Department of Physics \\
1150 University Avenue, Madison, WI 53706, USA}
\affiliation{$^{2}$ California Institute of Technology, Pasadena, CA, 91125 USA}
\date{\today}
\begin{abstract}
We discuss the simplest mechanisms for  generating neutrino masses  at tree level and one loop level. 
We find a significant number of new possibilities where one can generate neutrino masses at the one-loop level 
by adding only two new types of representations. These models have renormalizable interactions that automatically 
conserve baryon number. Adding to the minimal standard model a scalar color octet with 
$SU(3)\bigotimes SU(2) \bigotimes U(1)$ quantum numbers, $(8,2,1/2)$, and a fermionic color 
octet in the fundamental or adjoint representation of $SU(2)$ one can generate neutrino masses 
in agreement with  experiment. Signals at the LHC,  and constraints from flavour violation are briefly  discussed.  
\end{abstract}
\maketitle
\section{Introduction}
The existence of massive neutrinos is one of the main motivations for physics beyond the Standard Model (SM).
As is well-known the neutrinos can be Dirac or Majorana fermions. In the case of Majorana neutrinos 
there are a great variety of scenarios for the origin of neutrino masses.
At tree level we can generate neutrino masses using the well-known Type I, Type II or Type III seesaw scenarios. 
In the Type I seesaw mechanism one  adds at least two SM singlets, $\nu^C \sim (1,1,0)$~\cite{TypeI}, and once 
those singlets are integrated out the neutrino mass matrix is given by ${\cal M}^I = Y_\nu \  M_R^{-1} \ Y_\nu^T v^2$, 
where $Y_\nu$ is the Yukawa coupling between the SM leptonic doublet and the right-handed neutrinos, $v$ is the vacuum expectation value of the SM Higgs, and $M_R$ is the Majorana mass matrix for the right handed neutrinos. In the Type II seesaw mechanism~\cite{TypeII} an $SU(2)$ scalar triplet is introduced, $\Delta \sim (1,3,1)$, and the neutrino masss matrix reads as ${\cal M}^{II} =  h_\nu \ v_{\Delta}$. Here, $h_\nu$ is the Yukawa coupling between the leptons and the triplet, and $v_{\Delta}$ is the vacuum expectation value of the neutral component of the triplet. It is also possible to generate neutrino masses at tree level if one introduces at least two extra fermions in the adjoint representation of $SU(2)$, $\rho \sim (1,3,0)$~\cite{TypeIII},  and the mass 
matrix for neutrinos is similar to the Type I case, where one replaces $M_R$ by $M_\rho$,  the Majorana mass matrix for the fermionic triplets. This is the  Type III seesaw scenario. These are the simplest mechanisms for generating neutrino masses at tree level since they add just one new type of representation to the minimal Standard Model. If one realizes the Type III seesaw mechanism in the context of Grand Unified theories (GUT's)  it is always a hybrid scenario using  Type I plus 
Type III seesaw~\cite{TypeIII}.  Of course  one can add more than two fields as in the case of R-parity violation in SUSY theories.  See Ref.~\cite{Ma98} for the a review of different seesaw mechanisms. 

In this letter we investigate the simplest possible scenarios that generate the neutrino masses 
at one-loop level. We stick to the cases where there are at most  two new types of  fields with different  gauge
quantum numbers and restrict our attention, for the most part, to singlet, fundamental and adjoint 
representations of the non Abelian gauge groups.  We also focus on the case where the new particles have masses of order the ${\rm TeV}$ scale (or less) since then it may be possible to test the origin of neutrino masses experimentally.  

This letter is organized as follows: In the second section we discuss and classify 
the simplest scenarios for the generation of neutrino masses at one-loop level.
In Section III  phenomenological predictions and signals at the LHC 
are briefly discussed. We summarize our findings in the last section.
\section{Neutrino Masses: One-loop Mechanisms}
In this section we outline the simplest mechanisms where  neutrino masses are generated
at the one-loop level. It is well-known that introducing one scalar SM singlet field, 
$h \sim (1,1, 1)$ and an extra Higgs doublet, $(1,2,1/2)$, one can generate neutrino masses at the one loop-level. 
This is the so-called Zee model~\cite{Zee}.  In previous studies it has been shown that it 
is not possible to generate neutrino masses in agreement with the experiment~\cite{Koide} 
in the version of this model where  only one Higgs couples 
to the leptons (This naturally suppresses flavor changing neutral Higgs couplings.). This scenario is called the Zee-Wolfenstein model~\cite{Zee,Wolfenstein}.  
Introducing two scalar leptoquarks it is possible to generate neutrino masses 
at the one-loop level. One example  introduces the scalar fields 
$LQ_1 \sim (3,2,1/6)$ and $LQ_2 \sim (3,1,-1/3)$. However, these fields have renormalizable baryon 
number violating couplings  and proton decay occurs at tree level. A mechanism is needed to 
suppress baryon number violation if both fields are light. One 
can impose by hand some symmetries to forbid or suppress the proton decay rate or have the renormalizable coupling constants that violate baryon number be very small.  In this paper we restrict our attention to models where baryon number conservation is an automatic symmetry of the renormalizable couplings. See Ref.~\cite{LQs} for recent studies of 
models where one generates neutrino masses at one loop level using the leptoquark fields $LQ_{1,2}$.

If one introduces just two new types of fields, a scalar $S$ and the other a fermion $\rho$,  neutrino masses can be generated at 
one-loop level as shown in Fig.~1. Since one can generate neutrino masses at tree level using fermionic singlets, 
$\nu^C \sim (1,1,0)$, fermionic triplets, $\rho \sim (1,3,0)$, or scalar triplets, $\Delta \sim (1,3,1)$, we do not allow
these  representations. We are mainly interested in  
cases where colored fields play a role in the generation of neutrino masses since they can be produced at the LHC with large 
cross sections when their masses are below a TeV. 
\begin{center}
\begin{picture}(200,120)(0,0)
\ArrowLine(20,0)(60,0)
\ArrowLine(100,0)(60,0)
\ArrowLine(100,0)(140,0)
\ArrowLine(180,0)(140,0)
\DashArrowLine(68,72)(100,40)6
\DashArrowLine(132,72)(100,40)6
\DashArrowArc(100,0)(40,90,180)6
\DashArrowArcn(100,0)(40,90,0)6
\Text(40,-8)[c]{$\nu_i$}
\Text(80,-8)[c]{$\rho$}
\Text(100,0)[c]{$\times$}
\Text(120,-8)[c]{$\rho$}
\Text(160,-8)[c]{$\nu_j$}
\Text(59,28)[c]{$S$}
\Text(142,28)[c]{$S$}
\Text(60,80)[c]{$H^0$}
\Text(140,80)[c]{$H^0$}
\end{picture}
\vskip 0.2in
{\bf Fig.~1.} ~ Mechanism at one-loop level.
\end{center}
Let us now analyze the different scenarios where neutrino masses are generated through Fig.~1.

\begin{itemize}
\item \underline{Case 1)} In this case  two fermionic, $\chi \sim (1,2,0)$, fields and two extra 
scalars, $S \sim (1,3,1/2)$, are added to the minimal Standard Model. The new fields, $S$ and $\chi$ 
occur  inside the loop of Fig.~1. The simultaneous presence of the Yukawa interactions and the quartic 
interaction between $S$ and $H$ tell us that the lepton number is broken by two units generating 
the usual dimension five lepton-number violating operator for neutrino masses. Notice that in this case the extra 
fields do not have direct  couplings  to  the SM quarks. The interactions needed to realized this mechanism are:
\begin{equation}
-{\cal L}_1 = Y_1 \ l^T \ C \ i \sigma_2 \ S \ \chi \ + \ M_\chi \ \chi^T  \ C \ i \sigma_2 \  \chi 
\ + \ \lambda_1 \  H^T \ i \sigma_2 \  S^\dagger S^\dagger \ H \ + \ \rm{h.c.}  ~.
\label{V1}
\end{equation}
Unfortunately, since the extra fields $S$ and $\chi$ give rise to fractionally charged (color singlet) particles
there is always a stable charged particle in this scenario. Therefore  this case is ruled out by cosmological constraints and searches of exotic nuclei.   

\item \underline{Case 2)} One can have alternative mechanisms where the extra fields live in 
non-trivial representations of $SU(3)$. In order to avoid new anomalies we stick to 
real representations, of $SU(3)$, the one with lowest dimension being the adjoint.  Adding one (two) extra scalar $S_1 \sim (8,2,1/2)$ 
and  two (one) fermionic fields $\rho_1 \sim (8,1,0)$ it is possible to generate neutrino 
masses at one-loop level via Fig.~1.  Since the scalar octet has hypercharge $1/2$ one can use the quartic interactions 
between this field and the SM Higgs in order to generate the dimension five operator 
for neutrino masses. Notice that $\rho_1$ has the same quantum numbers as the gluino 
in supersymmetric models and $S_1$ will have extra couplings to the Standard Model quark
fields~\cite{Manohar}. As in the previous cases we show explicitly the relevant 
interactions (here we write just one possible quartic interaction for simplicity) needed to generate neutrino masses,
\begin{equation}
-{\cal L}_2 = Y_2 \ l^T \ C \ i \sigma_2 \ S_1 \  \rho_1 \ + \ M_{\rho_1} \ {\rm Tr} \ \rho^T_1 \ C \ \rho_1 
\ + \ \lambda_2 \  {\rm Tr} \left( S_1^\dagger \ H \right)^2 \ + \ {\rm h.c.} ~.
\label{V2}
\end{equation}
In Eq.~(\ref{V2}) the trace is over color matrices. In this case one can have a consistent scenario for cosmology since 
the scalar octet has couplings to the SM matter fields and 
we can satisfy all cosmological constraints. As far as we know this mechanism has not 
been discussed in previous studies.

\item \underline{Case 3)} One can generalize the previous mechanism using 
the extra scalar octet $S_1 \sim (8,2,1/2)$ and taking the extra fermion field in the adjoint 
representation of  $SU(2)$,  $\rho_2 \sim (8,3,0)$. One needs two (one) extra scalars and one (two) fermion in 
order to generate neutrino masses and mixings in agreement with  experiment.  In this case, in order to 
generate neutrino masses, one uses the following interactions:
\begin{equation}
-{\cal L}_3 = Y_3 \ l^T \ C \ i \sigma_2 \  \rho_2 \  S_1 \ + \ M_{\rho_2} \ {\rm Tr} \ \rho^T_2 \ C \ \rho_2 
\ + \ \lambda_3 \ {\rm Tr} \left( S_1^\dagger \  H \right)^2 \ + \ \rm{h.c.}~.  
\label{V3}
\end{equation}
As in the previous case one can satisfy all cosmological constraints since there are no stable charged particles and the
extra scalar octet has renormalizable Yukawa couplings to the Standard Model quarks.

\item \underline{Case 4)} It is also possible to introduce two copies of extra fermions which 
are in the adjoint representation of $SU(3)$ and in the fundamental of $SU(2)$, $\eta \sim (8,2,0)$, 
and two extra scalars  in the adjoint of both gauge groups, $\Sigma \sim (8,3,1/2)$.
Notice that in this case the extra scalar octets do not have Yukawa couplings to  the Standard Model quark 
fields since they are in the adjoint of $SU(2)$. Using the following interactions neutrino masses are generated at one loop
via Fig.~1
\begin{equation}
- {\cal L}_4 = Y_4 \ l^T \ C \ i \sigma_2 \  \Sigma \ \eta \ + \ M_\eta \ {\rm Tr} \ \eta^T \ C \ i \sigma_2 \  \eta
\ + \ \lambda_4 \ {\rm Tr} \ H^T \ i \sigma_2 \  \Sigma^\dagger \ \Sigma^\dagger \ H \ + \ \rm{h.c.} . 
\label{V4}
\end{equation}
However, this scenario is also ruled out since it has fractionally stable charged (color singlet) particles.

\end{itemize}
In Table 1 we summarize the different scenarios showing the  $SU(3)\bigotimes SU(2) \bigotimes U(1)$ gauge
quantum numbers of the different needed representations (The color representation $8$ can be replaced 
by any real representation $R$. Higher dimension $SU(2)$ representations are also possible.).
We have shown that to generate neutrino masses at one-loop level adding the minimal number of 
new representations, and imposing no extra symmetries, the most economical  ways are the Zee model and models that  introduce 
a scalar octet, $(8,2,1/2)$, and a fermionic octet which can be in the fundamental or adjoint representation of $SU(2)$.    
\begin{table}[tb]
\begin{center}
\begin{tabular}[t]{|c|c|c|c|}
  \hline
  \hline
   {\bf Seesaw Scenario} & {\bf Extra Scalar Representations} & {\bf Extra Fermion Representations}  & {\bf Status} \\
  \hline
  \hline
  {\bf Tree Level} & &  & \\
  \hline
     I   &  & (1,1,0)  & OK \\
     II  &  (1,3,1) &  & OK  \\
     III & & (1,3,0)  & OK \\
     \hline
     {\bf One Loop Level} & & & \\
     \hline
     Zee model & (1,1,1) &  & \rm{OK\footnote{The Zee-Wolfenstein model, where only one of the Higgs doublets couples to the leptons,  is  ruled out~\cite{Koide}.}} \\
     Case 1)  & (1,3,1/2) & (1,2,0)  & \rm{ruled out} \\
     Case 2)  & (8,2,1/2) & (8,1,0) & OK \\
     Case 3)  & (8,2,1/2) & (8,3,0)  & OK \\
     Case 4)  & (8,3,1/2) & (8,2,0)  & \rm{ruled out} \\     
  \hline
  \hline
\end{tabular}
\end{center}
\caption{Different Seesaw Scenarios} 
\label{int}
\end{table}

Consider case 2 with two copies of the new fermions. Working in the mass eigenstate 
basis for the two new fermions $\rho_1^{\alpha}$ the neutrino mass matrix reads as,
\begin{equation}
\label{mnuformula}
{\cal M}_\nu^{ij}=Y_{2}^{i \alpha} \ Y_2^{j \alpha} \  \frac{\lambda_2}{16 \pi^2} \  v^2 \ I \left( M_{\rho_1^\alpha}, M_{S_1}\right), 
\end{equation}
with $\alpha=1,2$. The loop integration factor,  $I \left( M_{\rho_1^\alpha}, M_{S_1}\right)$, is given by,
\begin{equation}
I \left( M_{\rho_1^\alpha}, M_{S_1}\right)=4 M_{\rho_1^\alpha} \left({M_{S_1}^2-M_{\rho_1^\alpha}^2+M_{\rho_1^\alpha}^2{\rm ln}(M_{\rho_1^\alpha}^2/M_{S_1}^2) \over  \left(M_{S_1}^2- M_{\rho_1^\alpha}^2\right)^2 } \right)
\end{equation}
 With just this minimal number of copies of the new fields there is a massless neutrino.
Therefore, there are two types of spectra: Normal Hierarchy with $m_1=0$, $m_2 = \sqrt{\Delta m_{sol}^2}$, and 
$m_3=\sqrt{\Delta m_{sol}^2 \ + \ \Delta m_{atm}^2}$, and Inverted Hierarchy with 
$m_3=0$, $m_2 = \sqrt{\Delta m_{atm}^2}$, and $m_1=\sqrt{\Delta m_{atm}^2 \ - \ \Delta m_{sol}^2}$.
Here $\Delta m_{sol}^2 \approx 8 \times 10^{-5} \ \rm{eV}^2$ and  
$\Delta m_{atm}^2 \approx 2.5 \times 10^{-3} \ \rm{eV}^2$ are the solar and atmosphere 
mass squared differences. We  can also have a minimal scenario 
with two extra octet scalars and one fermionic octet.  

In the limit  $M_{S_1} \gg M_{\rho_1}$ the neutrino mass matrix becomes,
\begin{equation}
\label{mnuformula1}
{\cal M}_\nu^{ij}=Y_{2}^{i \alpha} \ Y_2^{j \alpha} \  \frac{\lambda_2}{4 \pi^2} \  v^2 \ \frac{M_{\rho_1^\alpha}}{M_{S_1}^2}.
\end{equation}
Using as input parameters, $M_{\rho_1} = 200$ GeV, $v=246$ GeV and $M_{S_1}=2$ TeV we find that in order 
to get the neutrino ``scale", $\sim 1$ eV, the combination of the couplings, $Y_2^2 \lambda_2 \sim 10^{-8}$. 
If $\lambda_2 \sim 1$ the elements of theYukawa coupling matrix, $Y_2 \sim 10^{-4}$. The 
Yukawa couplings can be larger  if $\lambda_2$ is smaller. In the scenarios proposed by us it is possible to reproduce the measured neutrino masses and 
mixing using a mechanism that can be tested at the LHC.

As we have discussed before, the simultaneous presence of the Yukawa term proportional to $Y_2$ 
and the quartic interaction proportional to $\lambda_2$ in Eq.~(\ref{V2}) violate lepton number. 
In this case, lepton flavor conservation is violated, even when $\lambda_2=0$, by the Yukawa 
couplings $Y_2$.  Hence, even when $\lambda_2$ is very small, there are constraints on 
the size of these Yukawa coupling constants from limits on the rates for lepton flavor 
violating processes like $\mu \rightarrow e +\gamma$. We hope to investigate these 
constraints in a future publication.

\section{Phenomenological Aspects}
{\underline{Possible Signals}:} We now discuss a few of the phenomenological aspects of the 
scenarios discussed above assuming that the extra fields have masses  of order a TeV (or less)  
so that they can be produced at the LHC. In case 2  neutrino masses are generated at one-loop 
using  the octet scalar, $S_1 \sim (8,2,1/2)$, and the fermionic octet, $\rho_1 \sim (8,1,0)$. The phenomenological aspects of the scalar octet have been studied in great detail by several groups~\cite{Octet}. If a neutral scalar is the lightest new particle  it will decay  directly to quark- antiquark pairs at tree level or at the one loop level to gluons using the interaction term in the scalar potential, $\lambda_5 {\rm Tr }H^{\dagger}SSS^{\dagger} +{\rm h.c.}$~\cite{Manohar}. In case 2 the extra fermion has the same quantum numbers as the gluino in SUSY theories. It can be produced in pairs  through the strong interactions and its decay width will be dominated by two body decays, $\rho_1 \to \textit{l} \ S_1$ if  $M_{\rho_1} \ > \ M_{S_1} \ + \ M_{\textit l} $, or  three body 
decays when $S_1$ is virtual. Since the $\rho_1$ fields are Majorana  one can have very 
exotic final state channels with two sign-same charged leptons and four jets. 
In particular  the channels, 
$pp \ \to \ \rho_1 \rho_1 \ \to \ S_1^+ S_1^+  e^-_i   e^-_j \ \to \ e^-_i e^-_j t t \bar{b} \bar{b}$, 
where one has leptons with the same electric charge, two tops and two anti-bottom quarks is the cleanest channel to test the mechanism for neutrino masses in case 2.

From the production and decay properties  of the scalar octets one may be able to determine their masses, 
and this information can be used to understand the three or two body decays of the fermionic octet.  
Once we impose the constraints coming from neutrino masses one has some information 
on the Yukawa coupling between the octets and the leptonic doublets.  Unfortunately they are 
not completely determined since the coupling $\lambda_3$ multiplies the neutrino mass 
matrix (see Eq.~(\ref{mnuformula})). 

The phenomenology of case 3 is similar to case 2, 
however, one important difference is that there are both electrically charged and neutral 
color octet fermions. The splittings between the charged and neutral color octet fermions 
is small since it is generated at the one loop level. One striking channel associated with the production  
of  charged ``gluinos" at the LHC  is, $pp \ \to \ \rho_2^+ \rho_2^- \ \to \ e^+_i e_j^- S_1^0 S_1^0 \ \to \ e^+_i e^-_j t \bar{t} t \bar{t}$.

We can generalize cases 2 and 3 above by, for example, changing  the color octet representation 
to any other real representation $R$ of the SU(3) gauge group.  Interactions that break 
the $\rho \rightarrow -\rho$, $S \rightarrow -S$ symmetry are needed to allow for the 
new strongly interacting particles to decay. A term in the scalar potential of the form,  
$H^{\dagger}SSS^{\dagger}$,  breaks this discrete symmetry. It allows the neutral members 
of the $S$ representation to decay to  gluons at the one loop level. The octet representation is 
the smallest dimension real representation. It is also the only representation that allows for 
renormalizable  Yukawa couplings  of the scalars $S$ to the quarks. In the case  $R=8$  the $S$ 
scalars can decay at tree level to quark-antiquark pairs . 

{\underline{Other aspects}:} If one does not impose minimal flavor~\cite{mfv} violation the 
Yukawa couplings of the $S_1$ to quarks are constrained by the smallness of observed 
flavor changing neutral currents. For example, the measured value of the $K_L-K_S$ mass difference 
implies that the $d \rightarrow s$  Yukawa coupling of neutral $S$ to quarks  is less than about, 
$10^{-5}( M_{S_1}/{\rm TeV})$. This might seem like a very strong constraint, however, it is 
important to remember that the electron Yukawa  coupling of the standard model Higgs doublet 
is about $10^{-5}$. 

Before finishing we would like to comment on possible constraints 
coming from neutrinoless double beta decay. Cases 2 and 3 have the usual contribution 
to this rare process due to the existence of light Majorana neutrino masses. In addition 
the Yukawa couplings of the scalar octet to the quarks gives rise to new contributions. However, they  are highly suppressed by the masses of the scalar and fermionic octets.
\section{Summary}
We have discussed the simplest mechanisms that generate 
neutrino masses either at tree level or at one loop where one  introduces 
at most two types of  representations beyond those that are in the minimal Standard Model. We found new possibilities where  
neutrino masses are generated at the one-loop level  and the renormalizable interactions automatically conserve baryon number.
The simplest cases  have a scalar octet, 
$(8,2,1/2)$, and a fermionic octet in the fundamental or adjoint 
representation of $SU(2)$.   Possible signals at the LHC, 
and the constraints from flavour violation  were briefly  discussed.  We hope to elaborate on some of the phenomenological implications of  these models in a future publication. 
\subsection*{Acknowledgment}
One of the authors (P.F.P.) would like to thank Caltech for hospitality.
The work of P. F. P. was supported in part by the U.S. Department of Energy
contract No. DE-FG02-08ER41531 and in part by the Wisconsin Alumni
Research Foundation.  The work of M.B.W. was supported in part by the U.S. Department of Energy contract No. DE-FG02-92ER40171.


\end{document}